\documentclass[%
 reprint,
 amsmath,amssymb,
 aps,
]{revtex4-2}

\usepackage{graphicx}
\usepackage{dcolumn}
\usepackage{bm}

\begin{document}

\preprint{APS/123-QED}

\title{Systematic study of macroscopic interaction models and short-range effects in
$^{12}$C+$^{12}$C fusion}

\author{Azni Abdul Aziz}
 \email{Corresponding Author: azniabdulaziz@iium.edu.my}

\affiliation{%
 Department of Physics, Kulliyyah of Science, International Islamic University Malaysia, 25200, Kuantan, Pahang, Malaysia
}%

\author{Norhasliza Yusof}
 \affiliation{
 Department of Physics, Faculty of Science, University of Malaya, 50603 Kuala Lumpur, Malaysia\\
Center of Astronomy and Astrophysics, Faculty of Science, University of Malaya, 50603 Kuala Lumpur, Malaysia
}%

\date{\today}

\begin{abstract}
The \(^{12}\mathrm{C}+^{12}\mathrm{C}\) fusion reaction is central to stellar carbon burning, yet its cross section at astrophysical energies remains uncertain and theoretical extrapolations depend sensitively on the adopted nucleus–nucleus interaction. We systematically assess 24 macroscopic interactions within a common one-dimensional fusion framework and find that their relative agreement with experimental data depends strongly on both the statistical measure and energy interval considered, with no single interaction providing a uniformly superior description across all criteria. A common phenomenological short-range repulsive modification is then applied to all interactions without refitting their underlying parametrizations. Although the external fusion barrier remains essentially unchanged, the modification substantially reorganizes the inner-potential structure and produces strongly model-dependent changes in agreement with experiment. Detailed analysis of five representative interactions shows that the resulting low-energy suppression propagates into the thermonuclear reaction rates, but does not produce the conventional logarithmic-slope signature of fusion hindrance over the calculated energy range. These results demonstrate that additional short-range repulsion does not provide a universal improvement of macroscopic fusion potentials and that the reliability of extrapolations toward stellar carbon-burning energies remains sensitive to both the choice of interaction and its short-range behavior.
\end{abstract}

\maketitle
\setlength{\parskip}{0pt plus 1pt}
\raggedbottom

\section{Introduction}

The $^{12}$C+$^{12}$C fusion reaction is a key nuclear
process in stellar carbon burning and influences the subsequent
evolution of massive stars. It is also relevant to explosive
astrophysical environments, including thermonuclear supernovae
and carbon-fueled X-ray superbursts~\cite{Chieffi2025}.
Reliable knowledge of the fusion cross section at astrophysical
energies is therefore important for stellar-evolution and
nucleosynthesis calculations.

Experimentally, the reaction remains difficult to constrain at
stellar energies because the fusion cross section decreases
rapidly below the Coulomb barrier. Direct measurements are
challenged by increasingly small reaction yields and by
background contributions, although recent experiments have
extended the available data toward lower energies~\cite{Jiang2018,Tan2020,Fruet2020,Tan2024}.
The excitation function also exhibits pronounced resonance
structures whose possible persistence toward astrophysical
energies introduces an additional uncertainty in the reaction
rate~\cite{Spillane2007,Tumino2018,Jiang2018}. Consequently, extrapolation of the measured cross section into the stellar-energy region remains strongly dependent on assumptions concerning the unresolved low-energy behavior.

A related uncertainty is the possible occurrence of fusion hindrance, in which the fusion cross section decreases more rapidly at deep-sub-barrier energies than expected from conventional fusion descriptions. In heavier systems, this behavior has been associated with an increasing logarithmic slope of the excitation function and the possible appearance of a maximum in the astrophysical \(S\) factor. Jiang et al. extended these systematics to light heavy-ion reactions and suggested that a similar low-energy suppression could occur in \(^{12}\mathrm{C}+^{12}\mathrm{C}\) fusion~\cite{Jiang2007}. The evidence in carbon fusion remains inconclusive, however, because the relevant energy region is sparsely constrained and affected by pronounced resonance structure. More recently, Uzawa and Hagino showed that the inferred hindrance behavior depends sensitively on the functional form adopted to extrapolate the logarithmic slope toward lower energies~\cite{UzawaHagino2025}. Thus, the existence of a conventional fusion-hindrance signature in \(^{12}\mathrm{C}+^{12}\mathrm{C}\) remains unsettled, and suppression of the low-energy cross section alone does not establish the occurrence of hindrance.

Independent of the hindrance question, theoretical predictions at sub-barrier energies are sensitive to the adopted nucleus--nucleus interaction because relatively small differences in the radial potential can produce substantial changes in the tunneling probability~\cite{DuttPuri2010a}. Proximity-type interactions constitute a broad family of macroscopic potentials constructed from nuclear surface and geometrical properties, while the Bass and Winther families provide alternative phenomenological descriptions of the heavy-ion interaction. Their different parametrizations can lead to different barrier and post-barrier structures and, consequently, different fusion predictions~\cite{Blocki1977,DuttPuri2010a}. Although such interactions have been extensively employed in heavy-ion fusion studies, their relative performance for the light \(^{12}\mathrm{C}+^{12}\mathrm{C}\) system, particularly across different sub-barrier energy intervals, has not been established within a common fusion framework. A systematic comparison is therefore needed to quantify the interaction-model dependence that enters extrapolations toward astrophysical energies.

Additional model dependence can arise at small internuclear separations, where the nuclear densities strongly overlap and the inner part of the ion--ion interaction becomes increasingly important. Mi\c{s}icu and Esbensen proposed that nuclear-matter saturation and incompressibility generate additional short-range repulsion and demonstrated that incorporating such repulsion can produce a shallower inner pocket and a thicker tunneling region~\cite{MisicuEsbensen2006}. Related microscopic and phenomenological studies likewise indicate that overlap-region physics can modify the inner interaction. This motivates examining how macroscopic potentials with different underlying radial structures respond to the same additional short-range repulsion. Applying a common modification, rather than independently readjusting each interaction, provides a controlled test of whether such repulsion produces a systematic improvement or whether its effect depends on the underlying macroscopic potential. This test does not assume that low-energy suppression necessarily corresponds to conventional fusion hindrance.

In the present work, we systematically investigate \(^{12}\mathrm{C}+^{12}\mathrm{C}\) fusion using 24 macroscopic nucleus--nucleus interactions within a common one-dimensional barrier-penetration framework. Their calculated fusion cross sections are assessed against a common experimental database using complementary logarithmic statistical measures over the full measured range and in separate energy intervals. A common phenomenological short-range repulsive modification is subsequently applied to all 24 interactions without refitting their underlying parametrizations, allowing the baseline performance of each interaction to be distinguished from its response to an identical modification of the short-range potential.

Five representative interactions are then examined in
greater detail through their radial-potential structure,
fusion excitation functions, astrophysical factors,
logarithmic-slope behavior, and thermonuclear reaction
rates. Particular attention is given to whether the
suppression induced by the short-range modification is
accompanied by the conventional signature associated with
fusion hindrance and to how the resulting model dependence
propagates into the $^{12}$C+$^{12}$C reaction rate at
stellar carbon-burning temperatures. Rather than seeking a
single universally optimal macroscopic interaction, the
aim is to quantify the model dependence associated with
both the choice of interaction and its short-range
behavior.

\section{Theoretical framework}

\subsection{Macroscopic nucleus--nucleus potentials}
\label{sec:potentials}

To examine the model dependence of the predicted
$^{12}$C+$^{12}$C fusion cross section, we consider 24
macroscopic nucleus--nucleus potentials within a common fusion
framework. The ensemble includes several proximity and
proximity-type formulations together with the Bass, Winther, and
Ng\^{o} potentials, thereby sampling different prescriptions for
the nuclear geometry, surface properties, and radial dependence of
the interaction. The complete set of potentials and their
corresponding references are summarized in
Table~\ref{tab:potentials}.

The nuclear interaction in the conventional proximity
formalism is written as
\begin{equation}
V_N(r)=4\pi\gamma b\,\overline{R}\,\Phi(\xi),
\label{eq:prox}
\end{equation}
where $\gamma$ is the nuclear surface-energy coefficient, $b$ is
the surface-width parameter, $\overline{R}$ contains the geometrical
dependence on the interacting nuclei, and $\Phi(\xi)$ is the
universal function of the scaled surface separation
\cite{Blocki1977}. Different proximity formulations modify one or
more of these ingredients.

The largest subgroup consists of 13 parametrizations based on the
Prox.~77 formalism \cite{Blocki1977}. These retain the same
geometrical prescription and universal function but employ different
surface-energy coefficients,
\begin{equation}
\gamma =
\gamma_0
\left[
1-k_s
\left(
\frac{N-Z}{N+Z}
\right)^2
\right].
\label{eq:gamma}
\end{equation}
For $^{12}$C, $N=Z$, so that $\gamma=\gamma_0$ and the differences
among these 13 interactions arise directly from the adopted values
of $\gamma_0$. The corresponding parameter prescriptions are taken
from Refs.~\cite{Myers1966,Myers1967,MollerNix1976,Krappe1979,
MollerNix1981,RoyerRemaud1984,MollerNix1988,Moller1995,
Pomorski2003}.

The remaining proximity-type interactions include Prox.~00,
Prox.~00DP, Prox.~2010, and Dutt2011, which incorporate alternative
prescriptions for quantities such as the nuclear radii, surface
properties, and universal function
\cite{Myers2000,DuttPuri2010a,DuttBansal2010,Dutt2011}.
The ensemble is completed by the Bass73, Bass77, and Bass80
potentials \cite{Bass1973,Bass1977,Bass1980};
CW76, BW91, and AW95
\cite{Christensen1976,BrogliaWinther1991,Winther1995};
and Ng\^{o}80 \cite{Ngo1980}. 

\begin{table}
\caption{
Macroscopic nucleus--nucleus potentials considered in the present
$^{12}$C+$^{12}$C fusion analysis.
}
\label{tab:potentials}
\begin{ruledtabular}
\begin{tabular}{ll}
Potential & Reference \\
\hline
Prox.~77 Sets 1--13 & \cite{Ghodsi2016} \\
Prox.~00            & \cite{Myers2000} \\
Prox.~00DP          & \cite{DuttPuri2010a} \\
Prox.~2010          & \cite{DuttBansal2010} \\
Dutt2011            & \cite{Dutt2011} \\
Bass73              & \cite{Bass1973} \\
Bass77              & \cite{Bass1977} \\
Bass80              & \cite{Reisdorf1994} \\
CW76                 & \cite{Christensen1976} \\
BW91                 & \cite{BrogliaWinther1991} \\
AW95                 & \cite{Winther1995} \\
Ng\^{o}80            & \cite{Ngo1980}
\end{tabular}
\end{ruledtabular}
\end{table}

\subsection{Fusion cross-section calculations}
\label{sec:fusion_calculation}

For a center-of-mass energy $E$, the fusion cross section is
\begin{equation}
\sigma_{\mathrm{fus}}(E)
=
\frac{\pi\hbar^2}{2\mu E}
\sum_{\ell}(2\ell+1)T_{\ell}(E),
\label{eq:fusion_cs}
\end{equation}
\noindent where $\mu$ is the reduced mass and $T_{\ell}(E)$ is the
transmission probability for the effective potential
\begin{equation}
V_{\mathrm{eff}}(r,\ell)
=
V_C(r)+V_N(r)
+
\frac{\hbar^2\ell(\ell+1)}{2\mu r^2}.
\label{eq:effective_potential}
\end{equation}

\noindent Coulomb interaction was
described by
\begin{equation}
V_C(r)=\frac{Z_1Z_2e^2}{r}.
\label{eq:coulomb}
\end{equation}

The nuclear potential $V_N(r)$ was supplied externally and used without additional strength renormalization. Fusion is treated using the incoming-wave boundary condition (IWBC)
implemented in CCFULL \cite{Hagino1999} with one-dimensional barrier-penetrating limit. The same procedure was applied to every interaction without
model-dependent normalization or fitting to the fusion data.

\subsection{Statistical assessment}
\label{sec:statistical_analysis}

The calculated excitation functions were assessed against a
compilation of 390 experimental $^{12}$C+$^{12}$C fusion
cross-section measurements
\cite{Patterson1969,Mazarakis1973,High1977,Aguilera2006,
BarronPalos2006,Spillane2007,Zickefoose2011,Jiang2018,
Tan2020,Fruet2020,Tan2024}.

Because the fusion cross section varies by several orders of
magnitude, the comparison was performed in logarithmic space using
two complementary measures. The reduced logarithmic chi-square is
defined as
\begin{equation}
\chi^2_{\log}
=
\frac{1}{N}
\sum_{i=1}^{N}
\left[
\frac{
\log_{10}\sigma_{{\rm exp},i}
-
\log_{10}\sigma_{{\rm th},i}
}{
\delta\log_{10}\sigma_{{\rm exp},i}
}
\right]^2 ,
\label{eq:reduced_log_chi_square}
\end{equation}
where
\begin{equation}
\delta\log_{10}\sigma_{{\rm exp},i}
=
\frac{\delta\sigma_{{\rm exp},i}}
{\sigma_{{\rm exp},i}\ln 10}.
\label{eq:log_sigma_uncertainty}
\end{equation}
Here $N$ denotes the number of experimental data points with finite,
positive experimental uncertainties and corresponding theoretical
cross sections. 

The uncertainty-independent Log-RMSD is
\begin{equation}
{\rm Log\mbox{-}RMSD}
=
\left\{
\frac{1}{N}
\sum_{i=1}^{N}
\left[
\log_{10}
\left(
\frac{\sigma_{{\rm th},i}}
     {\sigma_{{\rm exp},i}}
\right)
\right]^2
\right\}^{1/2}.
\label{eq:log_rmsd}
\end{equation}
Thus, $\chi^2_{\log}$ accounts for the reported measurement
uncertainties, whereas Log-RMSD characterizes the overall
multiplicative deviation with equal weighting in logarithmic space.

Both measures were evaluated over the full experimental range and
in three nonoverlapping intervals:
$2\leq E_{\rm c.m.}<3$~MeV,
$3\leq E_{\rm c.m.}<5$~MeV, and
$5\leq E_{\rm c.m.}\leq8$~MeV.
For each range, the 24 interactions were ranked independently
according to $\chi^2_{\log}$ and Log-RMSD, with smaller values
indicating closer agreement with experiment. The same procedure was applied
to the original and short-range-modified interactions.

\subsection{Short-range repulsive modification}
\label{sec:short_range}

Macroscopic interactions are primarily constrained in the surface
and barrier regions, whereas additional many-body effects may become
important at smaller internuclear separations. In
$^{12}$C+$^{12}$C, microscopic calculations indicate that Pauli
effects can introduce additional inner repulsion, although this
alone does not produce a pronounced fusion-hindrance signature
\cite{Godbey2019}. Motivated by this sensitivity of the inner
interaction, we introduce a phenomenological short-range repulsive
term into the existing nuclear potenial.

The modified nuclear potential is
\begin{equation}
V_N^{\rm mod}(r)
=
V_N^{\rm ori}(r)+V_{\rm SR}(r),
\label{eq:modified_potential}
\end{equation}
with
\begin{equation}
V_{\rm SR}(r)
=
\frac{V_H}
{1+\exp[(r-R_H)/a_H]},
\label{eq:short_range}
\end{equation}
where $V_H$ and $a_H$ determine the strength and diffuseness of the
additional repulsion. Its radial position is associated with the
touching configuration,
\begin{equation}
R_H=R_{\rm touch}=R_1+R_2,
\label{eq:RH}
\end{equation}
where $R_1$ and $R_2$ follow the radius prescription of the
corresponding macroscopic interaction. Thus, the modification is
introduced at the model-dependent touching configuration rather than
at a common absolute radius.

For the systematic comparison, the same parameters
\begin{equation}
V_H=15~{\rm MeV},
\qquad
a_H=0.20~{\rm fm}
\label{eq:short_range_parameters}
\end{equation}
are applied to all 24 interactions without refitting to the fusion
data. Since $V_{\rm SR}(r)$ rapidly decreases for $r>R_H$, the
modification predominantly affects the inner interaction while
leaving the external barrier nearly unchanged.

The modified potentials were subjected to the same fusion and
statistical procedures as their original counterparts. For either
statistical measure $X$, the relative change is defined as
\begin{equation}
\Delta X(\%)
=
\frac{X_{\rm ori}-X_{\rm mod}}
     {X_{\rm ori}}
\times100,
\label{eq:percentage_change}
\end{equation}
such that $\Delta X>0$ denotes improved agreement with experiment
and $\Delta X<0$ denotes deterioration. Changes in statistical rank
are considered separately from changes in the absolute metric.

\section{RESULTS AND DISCUSSION}

\subsection{Statistical assessment of the original potentials}
\label{sec:original_statistics}

The statistical performance of the 24 original interactions is
summarized in Table~\ref{tab:original_statistics}. The rankings vary
substantially with both the statistical measure and the energy
interval, indicating that no single interaction provides a
universally optimal description of the measured excitation function.

\begin{table*}[t]
\caption{
Statistical assessment of the 24 original macroscopic interactions
against the experimental $^{12}$C+$^{12}$C fusion data. Results are
given for the full experimental range and for the three energy
intervals considered in the analysis. Numbers in parentheses denote
the rank of the interaction for the corresponding statistical
measure and energy interval.
}
\label{tab:original_statistics}
\begin{ruledtabular}
\begin{tabular}{lrrrrrrrr}
& \multicolumn{4}{c}{Reduced log-$\chi^2$, $\chi^2_{\log}$} &
  \multicolumn{4}{c}{Log-RMSD} \\
\cline{2-5}\cline{6-9}
Potential &
Full & 2--3 MeV & 3--5 MeV & 5--8 MeV &
Full & 2--3 MeV & 3--5 MeV & 5--8 MeV \\
\hline
Bass77 & 45.208 (1) & 12.795 (8) & 59.939 (2) & 29.756 (3) & 0.36181 (8) & 0.55394 (9) & 0.30583 (8) & 0.17829 (4) \\
Prox.~77 Set~3 & 46.479 (2) & 12.990 (9) & 62.274 (5) & 28.930 (1) & 0.36319 (9) & 0.55683 (10) & 0.30795 (9) & 0.17256 (2) \\
Bass80 & 46.971 (3) & 16.268 (15) & 57.456 (1) & 47.433 (12) & 0.41447 (15) & 0.60183 (17) & 0.36970 (15) & 0.20889 (14) \\
CW76 & 47.150 (4) & 17.649 (18) & 61.475 (4) & 30.283 (4) & 0.41943 (16) & 0.62080 (20) & 0.37221 (16) & 0.17637 (3) \\
BW91 & 47.501 (5) & 13.498 (12) & 63.653 (7) & 29.139 (2) & 0.37031 (10) & 0.56444 (12) & 0.31654 (11) & 0.17192 (1) \\
Prox.~2010 & 48.962 (6) & 18.023 (19) & 61.205 (3) & 43.361 (8) & 0.43069 (19) & 0.62469 (21) & 0.38730 (19) & 0.20075 (11) \\
Prox.~00DP & 49.023 (7) & 13.447 (11) & 62.632 (6) & 34.870 (6) & 0.37504 (12) & 0.56371 (11) & 0.31510 (10) & 0.20404 (12) \\
AW95 & 50.863 (8) & 20.013 (22) & 65.516 (8) & 34.767 (5) & 0.44673 (20) & 0.64975 (22) & 0.40316 (20) & 0.18681 (5) \\
Prox.~77 Set~6 & 66.831 (9) & 11.776 (1) & 91.919 (9) & 41.769 (7) & 0.34755 (1) & 0.53570 (5) & 0.28908 (1) & 0.18926 (6) \\
Prox.~77 Set~10 & 70.558 (10) & 11.784 (2) & 97.005 (10) & 45.279 (9) & 0.34832 (2) & 0.53521 (4) & 0.29005 (2) & 0.19417 (7) \\
Prox.~77 Set~7 & 71.048 (11) & 11.788 (3) & 97.670 (11) & 45.755 (10) & 0.34845 (3) & 0.53518 (3) & 0.29022 (3) & 0.19483 (8) \\
Prox.~77 Set~4 & 73.008 (12) & 11.805 (5) & 100.32 (12) & 47.678 (13) & 0.34906 (5) & 0.53512 (1) & 0.29099 (4) & 0.19748 (10) \\
Prox.~77 Set~9 & 71.164 (13) & 11.788 (4) & 97.827 (13) & 45.867 (11) & 0.34849 (4) & 0.53517 (2) & 0.29027 (5) & 0.19499 (9) \\
Dutt2011 & 74.991 (14) & 11.839 (6) & 102.17 (14) & 53.130 (14) & 0.35056 (6) & 0.53572 (6) & 0.29222 (6) & 0.20478 (13) \\
Prox.~77 Set~5 & 87.629 (15) & 12.104 (7) & 119.81 (15) & 63.066 (15) & 0.35627 (7) & 0.53727 (7) & 0.29988 (7) & 0.21781 (15) \\
Prox.~77 Set~11 & 112.71 (16) & 13.079 (10) & 152.49 (16) & 92.103 (16) & 0.37490 (11) & 0.54777 (8) & 0.32214 (12) & 0.25226 (16) \\
Prox.~77 Set~2 & 139.57 (17) & 14.550 (13) & 186.95 (17) & 124.86 (17) & 0.39900 (13) & 0.56486 (13) & 0.34996 (13) & 0.28653 (17) \\
Ng\^{o}80 & 142.20 (18) & 14.656 (14) & 188.01 (18) & 138.06 (18) & 0.40138 (14) & 0.56687 (14) & 0.35086 (14) & 0.29836 (18) \\
Prox.~77 Set~1 & 170.14 (19) & 16.626 (16) & 225.94 (19) & 162.58 (19) & 0.42882 (17) & 0.58912 (15) & 0.38335 (17) & 0.32174 (19) \\
Prox.~77 Set~8 & 170.14 (20) & 16.626 (17) & 225.94 (20) & 162.58 (20) & 0.42882 (18) & 0.58912 (16) & 0.38335 (18) & 0.32174 (20) \\
Prox.~77 Set~12 & 188.86 (21) & 18.063 (20) & 249.78 (21) & 185.60 (21) & 0.44763 (21) & 0.60563 (18) & 0.40402 (21) & 0.34155 (21) \\
Prox.~77 Set~13 & 192.78 (22) & 18.377 (21) & 254.77 (22) & 190.39 (22) & 0.45159 (22) & 0.60920 (19) & 0.40834 (22) & 0.34553 (22) \\
Prox.~00 & 739.81 (23) & 52.283 (23) & 882.95 (23) & 1135.31 (23) & 0.85004 (23) & 0.91332 (23) & 0.82514 (23) & 0.83389 (23) \\
Bass73 & 1959.96 (24) & 223.12 (24) & 2548.35 (24) & 2091.88 (24) & 1.7487 (24) & 1.8285 (24) & 1.8434 (24) & 1.1527 (24) \\
\end{tabular}
\end{ruledtabular}
\end{table*}

Over the full experimental range, Bass77 gives the smallest reduced
logarithmic chi-square, $\chi^2_{\log}=45.208$, followed closely
by Prox.~77 Set~3 and Bass80. In contrast, Prox.~77 Set~6 gives the
smallest full-range Log-RMSD, 0.3476. The different ordering reflects
the complementary weighting of the two measures:
$\chi^2_{\log}$ emphasizes measurements according to their
reported uncertainties, whereas Log-RMSD characterizes the overall
logarithmic deviation without uncertainty weighting.

The energy-resolved analysis further changes the relative
performance. In the deepest interval,
$2\leq E_{\rm c.m.}<3$~MeV, Prox.~77 Set~6 and several closely
related Prox.~77 parametrizations form the leading group. In the
$3$--$5$~MeV interval, Bass80 gives the smallest
$\chi^2_{\log}$, whereas the Log-RMSD ranking favors
Prox.~77 Set~6 and related Prox.~77 interactions. At
$5\leq E_{\rm c.m.}\leq8$~MeV, Prox.~77 Set~3 is particularly
competitive, ranking first according to
$\chi^2_{\log}$ and among the leading interactions according to
Log-RMSD. 

These changes show that global agreement alone does not determine
the reliability of an interaction at sub-barrier energies. The
redistribution of the rankings with energy reflects differences in
the predicted energy dependence of the fusion excitation function,
rather than simply an overall normalization difference among the
models. Accordingly, the two statistical rankings are retained
independently and provide the baseline for assessing the response to
the short-range repulsive modification.

\subsection{Systematic response to the short-range repulsive modification}
\label{sec:modification_response}

\noindent We next examine the response of all 24 interactions to the common
short-range repulsive modification introduced in Sec.~II\,D.
Table~\ref{tab:modification_response} shows the corresponding
changes in $\chi^2_{\log}$ and Log-RMSD. The percentage change
follows Eq.~(\ref{eq:percentage_change}), with positive values
indicating improved agreement with experiment.

\begin{table*}[t]
\caption{
Percentage changes in the reduced logarithmic chi-square and
Log-RMSD following the short-range repulsive modification for all
24 interactions. The percentage change is defined by
Eq.~(\ref{eq:percentage_change}), such that positive values indicate
improved agreement with experiment and negative values indicate
deterioration. Numbers in parentheses give the rank of the
\emph{modified} interaction for the corresponding statistical
measure and energy interval.
}
\label{tab:modification_response}
\begin{ruledtabular}
\begin{tabular}{lrrrrrrrr}
& \multicolumn{4}{c}{$\Delta\chi^2_{\log}$ (\%)} &
  \multicolumn{4}{c}{$\Delta{\rm Log\mbox{-}RMSD}$ (\%)} \\
\cline{2-5}\cline{6-9}
Potential &
Full & 2--3 MeV & 3--5 MeV & 5--8 MeV &
Full & 2--3 MeV & 3--5 MeV & 5--8 MeV \\
\hline
Bass77 & -1.93 (3) & +3.18 (7) & -2.89 (5) & -0.15 (1) & +2.13 (6) & +1.16 (7) & +3.04 (6) & +2.26 (1) \\
Prox.~77 Set~3 & +3.20 (1) & +3.53 (9) & +4.18 (3) & -6.25 (3) & +1.78 (8) & +1.27 (9) & +2.67 (9) & -2.47 (2) \\
Bass80 & +1.24 (4) & +24.65 (3) & +5.86 (1) & -27.46 (13) & +13.87 (9) & +9.42 (3) & +20.59 (3) & -4.78 (13) \\
CW76 & +1.02 (5) & +2.21 (15) & +1.15 (4) & -1.08 (2) & +0.94 (14) & +0.81 (15) & +1.14 (14) & -0.35 (3) \\
BW91 & -42.14 (7) & -6.15 (12) & -37.39 (7) & -109.19 (14) & -3.48 (12) & -2.46 (12) & -2.70 (11) & -25.28 (12) \\
Prox.~2010 & +7.09 (2) & +14.71 (13) & +10.39 (2) & -17.65 (10) & +6.57 (13) & +5.53 (13) & +8.34 (13) & -5.31 (11) \\
Prox.~00DP & -25.57 (6) & +13.79 (1) & -32.39 (6) & -3.25 (4) & +7.19 (1) & +5.60 (1) & +9.98 (1) & +0.99 (9) \\
AW95 & -96.30 (14) & -72.92 (20) & -103.86 (14) & -56.95 (11) & -33.36 (19) & -22.67 (20) & -42.90 (19) & -20.98 (14) \\
Prox.~77 Set~6 & -14.38 (8) & -3.07 (2) & -16.48 (8) & -0.19 (5) & -0.34 (2) & -0.38 (2) & -0.34 (2) & -0.15 (4) \\
Prox.~77 Set~10 & -17.56 (9) & -4.72 (4) & -19.76 (9) & -3.22 (6) & -1.25 (3) & -0.81 (4) & -1.82 (4) & -1.18 (5) \\
Prox.~77 Set~7 & -18.01 (10) & -4.95 (5) & -20.21 (10) & -3.69 (7) & -1.38 (4) & -0.87 (5) & -2.02 (5) & -1.35 (6) \\
Prox.~77 Set~4 & -19.64 (12) & -5.83 (8) & -21.87 (11) & -5.43 (9) & -1.89 (7) & -1.10 (8) & -2.84 (8) & -1.95 (8) \\
Prox.~77 Set~9 & -18.10 (11) & -5.00 (6) & -20.31 (12) & -3.79 (8) & -1.41 (5) & -0.88 (6) & -2.07 (7) & -1.39 (7) \\
Dutt2011 & -22.96 (13) & -7.73 (10) & -25.92 (13) & -4.56 (12) & -2.57 (10) & -1.55 (10) & -3.89 (10) & -1.79 (10) \\
Prox.~77 Set~5 & -31.18 (15) & -12.48 (11) & -33.25 (15) & -20.55 (15) & -6.00 (11) & -2.89 (11) & -9.21 (12) & -7.60 (15) \\
Prox.~77 Set~11 & -50.21 (16) & -23.94 (14) & -51.24 (16) & -49.19 (16) & -13.68 (15) & -6.07 (14) & -20.14 (15) & -18.98 (16) \\
Prox.~77 Set~2 & -71.50 (17) & -38.83 (16) & -70.38 (17) & -84.93 (17) & -22.68 (16) & -10.33 (16) & -31.45 (16) & -32.27 (17) \\
Ng\^{o}80 & -659.43 (23) & -3618.87 (24) & -667.69 (23) & -160.40 (18) & -387.94 (24) & -412.96 (24) & -404.09 (23) & -44.04 (18) \\
Prox.~77 Set~1 & -103.65 (18) & -71.36 (17) & -99.38 (18) & -134.27 (19) & -36.17 (17) & -19.93 (17) & -46.04 (17) & -49.43 (19) \\
Prox.~77 Set~8 & -103.65 (19) & -71.36 (18) & -99.38 (19) & -134.27 (20) & -36.17 (18) & -19.93 (18) & -46.04 (18) & -49.43 (20) \\
Prox.~77 Set~12 & -130.36 (21) & -98.71 (21) & -124.19 (21) & -170.84 (22) & -48.05 (21) & -30.32 (21) & -58.54 (21) & -60.42 (22) \\
Prox.~77 Set~13 & -138.27 (22) & -110.00 (22) & -132.04 (22) & -178.42 (23) & -51.62 (22) & -35.00 (22) & -61.54 (22) & -62.74 (23) \\
Prox.~00 & +49.02 (20) & +38.87 (19) & +44.88 (20) & +62.05 (21) & +27.67 (20) & +18.35 (19) & +28.60 (20) & +39.17 (21) \\
Bass73 & +0.24 (24) & +4.81 (23) & +1.27 (24) & -5.92 (24) & +1.16 (23) & +2.32 (23) & +1.08 (24) & -2.39 (24) \\
\end{tabular}
\end{ruledtabular}
\end{table*}

The full-range results reveal a strongly model-dependent response.
Only a subset of the interactions improves according to both
statistical measures, while others deteriorate or show opposite
trends between $\chi^2_{\log}$ and Log-RMSD. Prox.~2010 shows
a clear favorable response under both criteria, while Bass80 also
benefits substantially, particularly in Log-RMSD. Prox.~77 Set~3
shows a more moderate improvement. Conversely, several interactions
that perform competitively in their original form do not benefit
from the same additional repulsion. Thus, good baseline performance
and favorable response to the modification are distinct properties
of an interaction.

The response is also energy dependent. The modification generally
has a stronger influence in the lowest-energy interval, where the
fusion probability is most sensitive to changes in the inner
interaction. Nevertheless, improvement at $2$--$3$~MeV does not
necessarily persist in the $3$--$5$ or $5$--$8$~MeV intervals.
A favorable full-range change can therefore conceal different
responses within individual energy regions, and vice versa.

Changes in rank were considered separately from changes in the
absolute statistical measures. In particular, an interaction may
move upward in rank even when its own metric deteriorates if other
models deteriorate more strongly. Rank improvement alone is
therefore not interpreted as evidence of improved agreement.

Overall, the all-model comparison demonstrates that the same
short-range repulsive modification does not systematically improve
the macroscopic interactions. Its effect depends on the underlying
potential and on the energy region considered. This heterogeneous
response motivates the potential-structure analysis in the following
subsection.

\subsection{Potential characteristics and origin of the
model-dependent response}
\label{sec:potential_diagnostics}

To investigate the origin of the model-dependent response, the
original and modified interactions were examined in terms of the
barrier height $V_B$, barrier radius $R_B$, barrier curvature
$\hbar\omega_B$, and inner-potential structure.

\begin{figure}[t]
    \centering
    \includegraphics[width=\columnwidth]
    {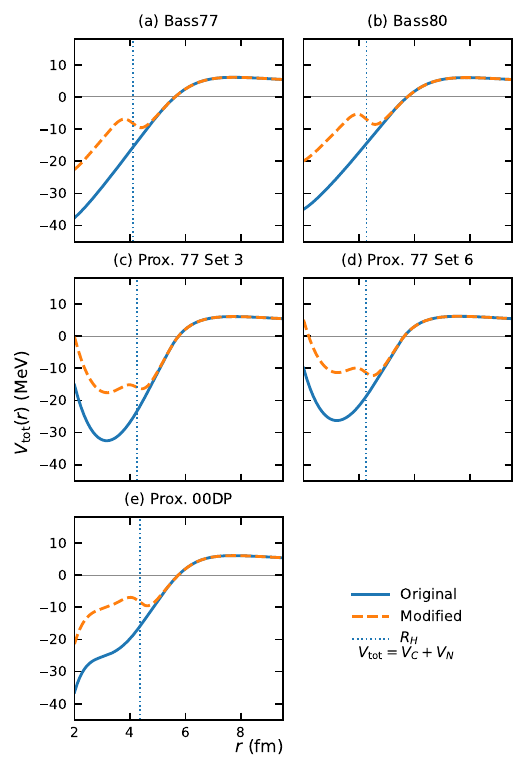}
    \caption{
    Total interaction potentials $V_{\rm tot}(r)=V_C(r)+V_N(r)$
    for the five representative interactions, comparing the original
    (solid curves) and short-range-modified (dashed curves)
    calculations. The vertical dotted lines indicate the corresponding
    overlap radius $R_H$. The modification leaves the external
    fusion-barrier region essentially unchanged while substantially
    restructuring the interaction at smaller internuclear separations.
    }
    \label{fig:potential_profiles}
\end{figure}

\begin{table*}[t]
\caption{
Barrier and inner-potential characteristics of the five representative
interactions. The listed barrier quantities refer to the original
interactions; their changes following the short-range repulsive
modification are negligible at the displayed precision.
$D_{\rm out}$ denotes the depth of the identified outer pocket
relative to the external barrier. A dash indicates that no distinct
outer pocket is identified by the numerical diagnostic. The final
column gives the number of identified inner minima before and after
modification.
}
\label{tab:potential_diagnostics}
\begin{ruledtabular}
\begin{tabular}{lcccccc}
Potential &
$V_B$ (MeV) &
$R_B$ (fm) &
$\hbar\omega_B$ (MeV) &
$D_{\rm out}^{\rm ori}$ (MeV) &
$D_{\rm out}^{\rm mod}$ (MeV) &
$N_{\min}^{\rm ori}\rightarrow N_{\min}^{\rm mod}$ \\
\hline
Bass77 & 6.1114 & 7.7391 & 2.6342 & 45.95 & 15.67 & $1\rightarrow2$ \\
Bass80 & 6.0247 & 7.8637 & 2.6111 & 42.19 & 14.57 & $1\rightarrow2$ \\
Prox.~77 Set~3 & 6.0629 & 7.7197 & 2.5430 & 38.61 & 22.46 & $1\rightarrow2$ \\
Prox.~77 Set~6 & 6.1541 & 7.5915 & 2.6028 & 32.43 & 18.44 & $1\rightarrow2$ \\
Prox.~00DP & 6.0819 & 7.7327 & 2.6061 & -- & 15.61 & $0\rightarrow1$ \\
\end{tabular}
\end{ruledtabular}
\end{table*}

A common feature of all 24 interactions is that the short-range
modification leaves the external fusion barrier essentially
unchanged. The changes in $V_B$ are at most of order
$10^{-2}$~keV, while the shifts in $R_B$ are of order
$10^{-4}$~fm or smaller; changes in $\hbar\omega_B$ are likewise
negligible. The modification therefore acts predominantly inside
the external barrier rather than altering the entrance-channel
barrier.

In contrast, the inner interaction is substantially reorganized.
Within the numerical diagnostic, 21 of the 24 interactions change
from one identified inner minimum to two minima after modification,
while two additional interactions develop an identified outer pocket
that is absent in their original form. Thus, nearly invariant barrier
properties can coexist with pronounced changes in the post-barrier
radial structure.

For the subsequent detailed analysis, Bass77, Bass80, Prox77 Set~3, Prox77 Set~6, and Prox00DP were chosen to represent the different performance characteristics observed across the considered energy intervals. The selection was guided by their complementary performance according to the reduced log-$\chi^2$ and Log-RMSD metrics over both the full energy range and the $2$--$3$, $3$--$5$, and $5$--$8$~MeV intervals. Bass77 provides strong overall agreement according to the full-range reduced log-$\chi^2$, while Prox77 Set~6 is distinguished by its low Log-RMSD and strong performance in the deep-sub-barrier region. Bass80 performs particularly well in the $3$--$5$~MeV interval, whereas Prox77 Set~3 provides consistently competitive behavior, especially toward the higher-energy region. Prox00DP was retained to broaden the representative set by including a potential with a contrasting statistical performance and response to the short-range modification. Thus, these five interactions span different energy-dependent behaviors and sensitivities to the statistical metric, providing a more informative basis for examining the effects of the short-range repulsive modification than a selection based solely on a single global ranking. Their barrier and pocket
characteristics are summarized in
Table~\ref{tab:potential_diagnostics}, while
Fig.~\ref{fig:potential_profiles} compares their original and
modified radial potentials.

For Bass77, Bass80, and the two Prox.~77 interactions, the
modification substantially reduces the outer-pocket depth and
introduces an additional inner minimum while leaving the external
barrier practically unchanged. Prox.~00DP behaves differently:
no distinct outer pocket is identified in its original form, whereas
the modified interaction develops a finite pocket near
$R=4.62$~fm with a depth of about 15.6~MeV.

The magnitude of the structural change does not, however, determine
the statistical response. Bass80 undergoes substantial inner
restructuring and benefits strongly from the modification, whereas
Bass77 exhibits an even larger reduction in pocket depth but only a
modest statistical benefit. Conversely, Prox.~77 Set~6 develops
clear additional inner structure while its overall statistical
agreement deteriorates. Pocket-depth reduction alone is therefore
not a predictor of improved fusion behavior.

The heterogeneous response is instead associated with the interplay
between the added repulsion and the pre-existing inner structure of
each interaction. Differences in pocket depth, the radial positions
of the inner extrema, and the overall post-barrier potential shape
modify the sub-barrier transmission differently even when the same
short-range prescription is applied. The consequences for the fusion
excitation functions are examined next.

\subsection{Fusion cross sections and astrophysical S-factor}
\label{sec:cross_section_sfactor}

\begin{figure}[t]
    \centering
    \includegraphics[width=\columnwidth]
    {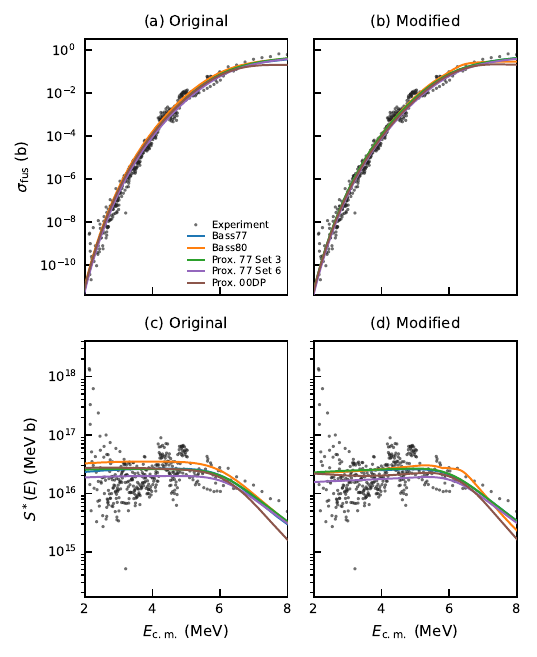}
    \caption{
    Fusion cross sections (upper panels) and modified astrophysical
    $S^*(E)$ factors (lower panels) for the five representative
    interactions. Results obtained with the original interactions are
    shown in panels (a) and (c), while those following the
    short-range repulsive modification are shown in panels (b) and
    (d). The symbols denote the compiled experimental
    $^{12}$C+$^{12}$C measurements
    \cite{Patterson1969,Mazarakis1973,High1977,Aguilera2006,
    BarronPalos2006,Spillane2007,Zickefoose2011,Jiang2018,
    Tan2020,Fruet2020,Tan2024}. The legend in panel (a) applies to all four panels.
    }
    \label{fig:cross_section_sstar}
\end{figure}

Figure~\ref{fig:cross_section_sstar}(a) compares the calculated
fusion cross sections for the five representative original
interactions with the experimental $^{12}$C+$^{12}$C data, while
Fig.~\ref{fig:cross_section_sstar}(b) shows the corresponding
short-range-modified calculations. The calculations are intended to
describe the smooth barrier-penetration component of the excitation
function rather than its detailed resonance structure.

The short-range modification predominantly suppresses the
sub-barrier cross section, while the original and modified
predictions converge toward higher energies. This behavior is
consistent with the potential diagnostics of Sec.~III\,C, which
show that the modification substantially changes the inner
interaction while leaving the external barrier nearly unchanged.
The magnitude of the suppression is nevertheless strongly model
dependent. Bass80 shows a pronounced favorable response, while
Prox.~77 Set~3 improves more moderately. In contrast, the additional
suppression of Prox.~77 Set~6 does not improve its overall
statistical agreement. Thus, stronger low-energy suppression does
not necessarily imply a better description of the measured
excitation function.

The corresponding modified astrophysical factor is defined as
\begin{equation}
S^*(E)
=
E\sigma(E)
\exp\left(
\frac{87.21}{\sqrt{E}}+0.46E
\right),
\label{eq:sstar}
\end{equation}
with $E$ in MeV and $\sigma(E)$ in barns. This representation,
commonly used for $^{12}$C+$^{12}$C fusion, reduces the dominant
Coulomb-energy dependence and makes differences among the calculated
excitation functions more apparent
\cite{Patterson1969,Tan2024}.
The corresponding modified astrophysical factors are shown in
Figs.~\ref{fig:cross_section_sstar}(c) and
\ref{fig:cross_section_sstar}(d) for the original and modified
interactions, respectively. The short-range modification can also shift or generate
local structures in $S^*(E)$, reflecting changes in the energy
dependence of the calculated fusion probability.

\subsection{Logarithmic-slope analysis and fusion hindrance}
\label{sec:log_slope}

Fusion hindrance is further examined through the logarithmic slope
of the energy-weighted fusion cross section,
\begin{equation}
L(E)
=
\frac{d}{dE}\ln[E\sigma(E)].
\label{eq:log_slope}
\end{equation}

\begin{figure}[t]
    \centering
    \includegraphics[width=\columnwidth]
    {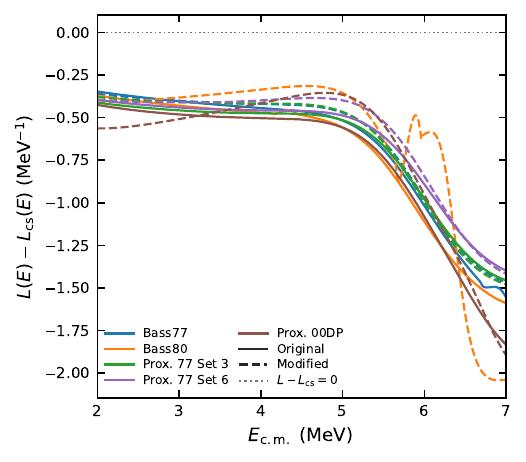}
    \caption{
    Difference between the logarithmic slope $L(E)$ and the
    constant-$S$ reference slope $L_{\rm cs}(E)$ for the five
    representative interactions. Solid and dashed curves denote the
    original and short-range-modified interactions, respectively.
    The horizontal dotted line corresponds to
    $L(E)-L_{\rm cs}(E)=0$; a crossing would indicate an extremum of
    the conventional astrophysical $S(E)$ factor. No crossing is
    obtained over the calculated energy range for either the original
    or modified interactions.
    }
    \label{fig:log_slope}
\end{figure}

For comparison, the logarithmic slope corresponding to a constant
conventional astrophysical $S$ factor is
\begin{equation}
L_{\rm cs}(E)=\frac{\pi\eta}{E},
\label{eq:constant_slope}
\end{equation}
where $\eta$ is the Sommerfeld parameter. A crossing between
$L(E)$ and $L_{\rm cs}(E)$ corresponds to an extremum of the
conventional $S(E)$ factor and is commonly used as a signature of
the onset of fusion hindrance \cite{Jiang2007}.

Figure~\ref{fig:log_slope} compares $L(E)$ and $L_{\rm cs}(E)$
for the five representative interactions. Although the short-range
repulsive modification changes the magnitude and energy dependence
of $L(E)$, none of the original or modified interactions exhibits
an $L(E)$--$L_{\rm cs}(E)$ crossing over the calculated energy
range. Thus, the additional low-energy suppression introduced by the
modification does not produce the conventional logarithmic-slope
signature of fusion hindrance.

This result is consistent with the continuing uncertainty regarding
hindrance in $^{12}$C+$^{12}$C fusion. Jiang \textit{et al.}
predicted a crossing by extrapolating the experimental logarithmic
slope \cite{Jiang2007}, whereas the recent reanalysis by Uzawa and
Hagino showed that the crossing disappears when the assumed energy
dependence of the extrapolation is relaxed
\cite{UzawaHagino2025}. Microscopic calculations have likewise
found no pronounced low-energy hindrance signature
\cite{Godbey2019}.

Within the present macroscopic framework, low-energy suppression and
fusion hindrance should therefore be distinguished. Additional
short-range repulsion can reduce the calculated fusion probability
and improve agreement with experiment for selected interactions
without generating a hindrance threshold according to the
conventional logarithmic-slope criterion.

\subsection{Thermonuclear reaction rates and astrophysical implications}
\label{sec:reaction_rates}

The model dependence of the sub-barrier fusion cross section
propagates directly into the thermonuclear reaction rate. The rate
per mole is calculated as
\begin{equation}
N_A\langle\sigma v\rangle
=
\frac{3.7318\times10^{10}}
{\sqrt{\mu}\,T_9^{3/2}}
\int
\sigma(E)E
\exp\left(-\frac{11.605E}{T_9}\right)dE ,
\label{eq:reaction_rate}
\end{equation}
where $E$ is in MeV, $\sigma(E)$ is in barns, and $\mu=6$~amu
for $^{12}$C+$^{12}$C. The integral was evaluated numerically over
the calculated cross-section grid. 

\begin{figure}[t]
    \centering
    \includegraphics[width=\columnwidth]
    {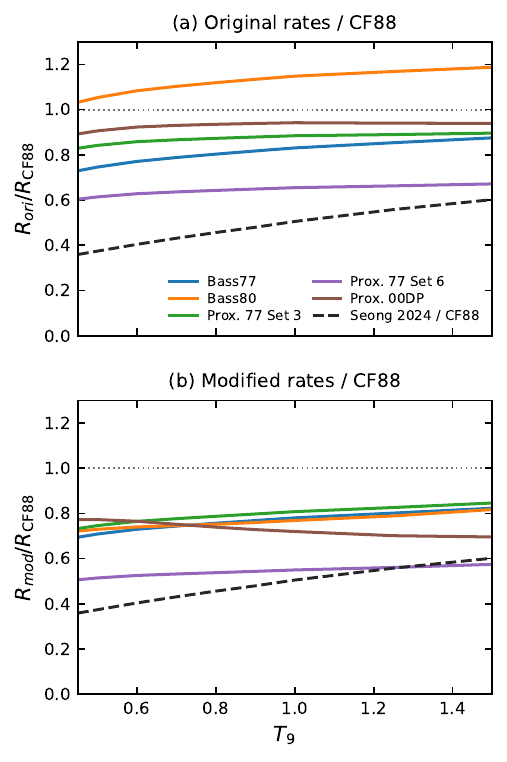}
    \caption{
    Thermonuclear $^{12}$C+$^{12}$C reaction rates relative to
    CF88 in the carbon-burning temperature range.
    Panel (a) shows the rates obtained from the five representative
    original interactions normalized to CF88, while panel (b) shows
    the corresponding short-range-modified rates.
    The dashed curve gives the Seong 2024 evaluation normalized to
    CF88, and the horizontal dotted line denotes unity.
    }
    \label{fig:rates_over_CF88}
\end{figure}

\begin{figure}[t]
    \centering
    \includegraphics[width=\columnwidth]
    {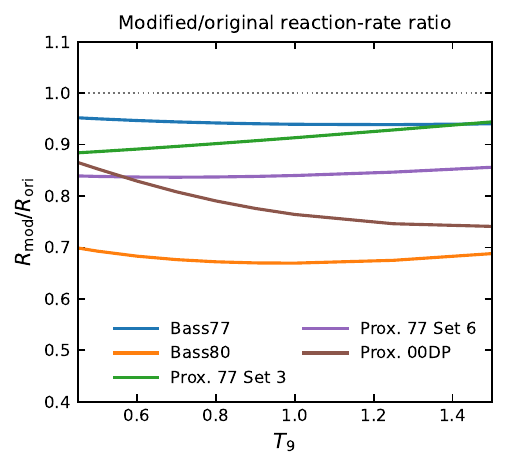}
    \caption{
    Modified-to-original thermonuclear reaction-rate ratios for the
    five representative interactions. The horizontal dotted line at
    unity corresponds to no change in the reaction rate. The
    short-range repulsive modification suppresses the rate throughout
    the displayed temperature interval, with a strongly
    model-dependent magnitude.
    }
    \label{fig:rate_mod_ori}
\end{figure}

Figure~\ref{fig:rates_over_CF88} compares the original and
short-range-modified reaction rates with the standard reaction rates of the
carbon fusion used in stellar evolution simulation provided
by Caughlan \& Fowler (1988) (CF88 hereafter).
~\cite{CaughlanFowler1988} and the recent reaction rate evaluated by Seong \cite{Seong2024}. Seong updated the carbon
fusion reaction rates by fitting the astrophysical S-factor data obtained from direct measurements based on the Fowler, Caughlan, \& Zimmerman (1975) formula~\cite{Fowler1975}. 

The normalization to CF88 makes the model spread and
the change produced by the modification directly visible. The short-range modification suppresses the
rate for all five interactions in the carbon-burning temperature
range, but by strongly model-dependent amounts. At $T_9=1$, the
modified-to-original ratios are approximately 0.94, 0.67, 0.91,
0.84, and 0.76 for Bass77, Bass80, Prox.~77 Set~3, Prox.~77
Set~6, and Prox.~00DP, respectively. Thus, the same short-range
modification changes the predicted rate by only about $6\%$ for
Bass77 but by about $33\%$ for Bass80.

The absolute rate also remains interaction dependent after
modification. At $T_9=1$, the five modified predictions span
approximately
$(2.12$--$3.12)\times10^{-11}$~cm$^3$~mol$^{-1}$~s$^{-1}$,
corresponding to about $0.55$--$0.81$ of CF88. The calculated
rates are broadly comparable in magnitude to the recent Seong evaluation. The modification-induced change is isolated in
Fig.~\ref{fig:rate_mod_ori}, which shows
$R_{\rm mod}/R_{\rm ori}$ for the five representative
interactions.

The effect of the short-range modification is also temperature
dependent rather than a constant rescaling of the rate. As $T_9$
changes, the Maxwellian integral samples different regions of the
fusion excitation function; consequently, the model-dependent
changes in sub-barrier transmission propagate into a
temperature-dependent rate uncertainty. This behavior is relevant
to stellar carbon burning, for which previous stellar-evolution
studies have shown that changes in the $^{12}$C+$^{12}$C rate can
affect carbon ignition and the subsequent burning evolution
\cite{Monpribat2022, Seong2024}.

The resulting rate variations demonstrate how uncertainty in the
low-energy nucleus--nucleus interaction can propagate into the
nuclear input used in stellar carbon-burning calculations.

\section{Summary and conclusions}
\label{sec:conclusions}

We have systematically investigated $^{12}$C+$^{12}$C fusion using
24 macroscopic nucleus--nucleus interactions within a common
one-dimensional barrier-penetration framework. Comparison with the
experimental excitation function using the reduced logarithmic
chi-square and Log-RMSD shows that the relative performance of the
interactions depends on both the statistical measure and the energy
range considered. No single potential provides a uniformly optimal
description across all criteria, emphasizing that a global ranking
alone does not characterize the reliability of a macroscopic
interaction at deep-sub-barrier energies.

Applying the same phenomenological short-range repulsive modification
to all 24 interactions reveals a similarly non-universal response.
Some potentials improve, others deteriorate, and the response can
change with energy. Potential diagnostics show that the modification
leaves the external fusion barrier essentially unchanged while
substantially reorganizing the inner interaction. However, the
magnitude of this restructuring does not correlate uniquely with the
statistical improvement. The effect of additional short-range
repulsion therefore depends on the pre-existing radial structure of
the underlying interaction rather than providing a universal
improvement of the fusion description.

For the five representative interactions examined in detail, the
modification produces model-dependent suppression of the
sub-barrier fusion cross section and corresponding changes in
$S^*(E)$. Nevertheless, none of the original or modified
calculations exhibits a crossing between $L(E)$ and the constant-$S$
reference slope $L_{\rm cs}(E)$ over the calculated energy range.
Within the present framework, additional low-energy suppression
therefore does not by itself constitute the conventional signature
of fusion hindrance.

The interaction dependence also propagates into the thermonuclear
reaction rate. Around the stellar carbon-burning regime, the common
short-range modification changes the rates of the representative
interactions by amounts ranging from several percent to approximately
one third, while a residual spread remains among the modified
predictions. These variations represent one component of the broader
uncertainty in the $^{12}$C+$^{12}$C rate associated with its
unmeasured low-energy behavior. Further experimental constraints at
deep-sub-barrier energies, together with microscopic guidance on the
short-range interaction, will therefore be important for reducing
model dependence in extrapolations of carbon fusion toward
astrophysical energies.

\begin{acknowledgments}
NY acknowledge the Fundamental Research Grant Scheme grant number FRGS/1/2018/STG02/UM/01/2 under Ministry of Higher Education, Malaysia
\end{acknowledgments}

\bibliography{Systematic_study}

\end{document}